\documentclass[%
 article.cls,
 reprint,
 numerical,
superscriptaddress,
showpacs,
 amsmath,amssymb,
 aps,prl,
 twocolumn,
floatfix,
nofootinbib,
]{revtex4-1}

\usepackage{graphicx}
\usepackage{dcolumn}
\usepackage{bm}
\usepackage{hyperref}

\usepackage{amssymb,epsfig}
\usepackage{amsmath}
\usepackage{epsfig}
\usepackage{color}
\usepackage{textcomp}
\usepackage[utf8]{inputenc}
\usepackage{natbib}
\usepackage[english]{babel}
\usepackage{commath}
\makeatletter
\newcommand{\newsec}[1]{\vspace{0.25cm}\noindent{\bf \emph{#1}}.\hspace{0.25cm}}
 
\newcommand{\vecr}{{\bf r}}
\newcommand{\HI}{{\rm HI}}
\newcommand{\s}{{\rm s}}
\newcommand{\g}{{\rm g}}
\newcommand{\quotes}[1]{``#1''}

\DeclareRobustCommand{\urlfootnote}{\hyper@normalise\urlfootnote@}
\makeatother 

\begin{document}


\title{Impact of  inhomogeneous CMB heating of gas on the HI 21-cm signal during dark ages }

\author{Sioree Ansar}
\email{sioreeansar@gmail.com}
\author{Kanan K. Datta}%
 \email{kanan.physics@presiuniv.ac.in}
\affiliation{%
 Department of Physics, Presidency University, 86/1 College Street, Kolkata-700073, India
}%


\author{Dhruba Dutta Chowdhury}
\affiliation{
Department of Astronomy, Yale University, New Haven, CT 06511
 \\
}%


\date{\today}

\begin{abstract}
Observations of redshifted 21-cm signal from neutral hydrogen (HI) appear to be the most promising probe of the cosmic dark ages. The signal carries information about the thermal state along with density distribution of the intergalactic medium (IGM). The cosmic microwave background radiation (CMBR), through its interaction with charged particles, plays a major role in determining the kinetic and spin temperature of HI gas in the IGM during dark ages. A Spatially fluctuating ionization fraction, which is caused by inhomogeneous recombinations, causes heat transfer from the CMBR to the IGM  gas inhomogeneous. We revisit the impact of  this inhomogeneous heat transfer on spatial fluctuations in the observed HI 21-cm signal over a large redshift range during dark ages. Our study shows that the effect negatively impacts fluctuations in the HI spin temperature and results in an enhanced HI 21-cm power spectrum. We find that the effect is particularly important during the transition of the gas kinetic temperature being coupled to the CMBR to fully decoupled from it, i.e., in the redshift range $30 \lesssim z \lesssim 300$. It is found that, on the average the HI power spectrum, $P_{T_b}(k, z)$ is enhanced by $\sim 4\%$, $\sim10 \%$ , $\sim 20\%$ and $\sim 30 \%$ at redshifts $60$, $90$, $140$, and $200$ respectively at $k=0.1 \, {\rm Mpc}^{-1}$. The effect becomes even more significant at lower values of $k_{\parallel}^2/k^2$ due to the reduced dominance of the peculiar velocity. It is observed that the power spectrum is enhanced by $\sim 49\%$ and $\sim 93\%$ at redshifts $140$ and $200$ respectively at  $k=0.1 \, {\rm Mpc}^{-1}$ for $k_{\parallel}^2/k^2=0$. This enhancement has a weak $k$-mode dependence. 



\end{abstract}

\pacs{Valid PACS appear here}
\maketitle


\newsec{Introduction}
\label{sec:intro}
The Universe started becoming neutral during the cosmic recombination epoch. Subsequently, it passed through a long dark episode, known as the cosmic dark ages, when there was no source of visible light. Although an important event, this epoch could not be explored yet due to lack of  observations. Observations using redshifted 21-cm signal from neutral hydrogen (HI) seem to be the only viable probe that has the potential to test and verify our current understanding of the dark ages by allowing the cosmological investigation of large scale matter density distribution and thermal history during the epoch\citep{2004PhRvL..92u1301L,2004Bharadwaj}.  This can further be used to address some of the key cosmological questions regarding the formation of the first gravitationally bound objects\citep{2007naoz}, origin and formation of seed black holes\citep{2013tashiro}, primordial non-Gaussianity\cite{2015munoz}, dark matter decay and annihilation\citep{2006Furlanetto,2009natarajan}, effect of primordial magnetic fields\citep{2009Schleicher}, variations in the fine structure constant\citep{2007Khatri}  etc. 

In the well-established standard cosmological scenario, the evolution of the background Universe and relevant physical processes that determine the strength of the HI 21-cm signal during dark ages are known with a high degree of accuracy. Further, the signal is not complicated by astrophysical processes unlike that during cosmic dawn and reionization epochs. This makes it possible to calculate the observable quantities i.e, the mean differential brightness temperature of the HI 21-cm signal and its spatial fluctuations, which can be well approximated as linear up to very small scales, with a high level of precision. 

A significant amount of work has already been done in order to understand both the globally averaged HI 21-cm signal and its spatial fluctuations during dark ages. Physical processes such as the recombination\citep{1968Peebles}, coupling of the gas kinetic temperature with the background CMBR temperature, collisions between HI and other species\citep{2005Zygelman}, radiative coupling of the HI spin temperature with the background CMBR temperature, that ultimately determine the redshift evolution of the HI 21-cm signal, have been studied extensively (see Ref. \citep{2012Pritchard,2006bFurlanetto,2018Pratika} for a review).

 This article focuses on the spatial fluctuations in the HI 21-cm differential brightness temperature. 
In particular, we highlight the role of inhomogeneous heat transfer from the CMBR to the IGM as an important contributor to the fluctuations. The fluctuations in the ionization fraction make the heating of gas inhomogeneous. This contributes to  the fluctuations in the gas kinetic temperature and, consequently, in the spin temperature. The effect has been included earlier \citep{2007Lewis,Lewis2007,2014Haimoud}. However, the focus was only at redshift $\sim 50$  where the effect was found to be  negligible on the large scale HI 21-cm power spectrum. Here, we investigate and quantify, in detail,  the impact of the effect  on the HI 21-cm power spectrum over a large redshift range  $30 \lesssim z \lesssim 300$. We adopt a simple analytical approach which clearly explains roles of various physical processes leading to the effect. We use  cosmological parameters ($\Omega_{\rm m0}$, $\Omega_{\Lambda 0}$, $\Omega_{\rm b0} h ^{2}$, h)=(0.3, 0.7, 0.02, 0.7), consistent with recent Planck measurements.  





\newsec{Basic 21-cm signal and evolution of mean quantities}
\label{sec:sec2}
The differential brightness temperature of HI 21-cm spin flip transition  with respect to the CMBR temperature  at  redshift $z$ can be written as\citep{2006bFurlanetto}
\begin{equation} \label{tb}
\begin{aligned}
T_{b} \approx {\bar T} (1+\delta_{b}) \left( \frac{T_{\s}-T_{\gamma}}{T_{s}} \right) \left ( \frac{H(z)}{1+z} \frac{1}{dv_{\parallel}/ dr_{\parallel}}\right),
\end{aligned}
\end{equation}
where ${\bar T} = 27\textrm{ mK } x_{\HI} \left( \frac{\Omega_{b0}h^2}{0.023} \right)\left( \frac{0.15}{\Omega_{m0}h^2 }\frac{1+z}{10} \right) ^{0.5}$. $x_{\HI}$, $\delta_{b}$ and $T_{\gamma}$ are the hydrogen neutral fraction,  fractional over density in baryons and background CMBR brightness temperature respectively. The term $dv_{\parallel}/dr_{\parallel}$ is the gradient of proper velocity of HI gas along the line of sight. The spin temperature $T_{\s}$ is defined by the relation $\frac{n_{1}}{n_{0}}=\frac{g_{1}}{g_{0}}e^{-T_{*}/T_{\s}}$, where $n_{1}$ and $n_{0}$ are the number densities of ground state HI in triplet and singlet states respectively and $g_{1}$ and $g_{0}$ are the respective degeneracies with $\frac{g_{1}}{g_{0}}=3$. $T_{*}=\frac{h_{p}\nu_{21cm}}{k_{B}}=0.068$ K. During the dark ages, $T_{s}$  is mainly governed by two processes: i) collisional coupling  with the gas kinetic temperature $T_{\g}$, and ii) radiative coupling with background CMBR. 
Therefore,  the spin temperature during this epoch can be written as\citep{1958Field}
\begin{equation} \label{ts}
T_{\s}^{-1} = \left(\frac{x_{\rm c}T_{\g}^{-1}+T_{\gamma}^{-1}}{1+x_{\rm c} }\right),
\end{equation}
where $x_{\rm c}=T_{*}C_{10}/A_{10}T_{\gamma}$ is the collisional coupling coefficient. $A_{10}$ is the Einstein coefficient for spontaneous emission.   $C_{10}=K_{\mathrm 10}^{\rm HH}n_{\rm HI}$ is the collisional deexcitation rate of HI from the triplet to singlet state due to HI-HI collisions.  $n_{\rm HI}$ is the number density of the neutral hydrogen atom. We do not consider the HI-electron and HI-proton collisions as they are expected to play negligible roles in the redshift range of our interest.  We use the fitting formula $K_{10}^{\rm HH}=3.1\times 10^{-17} T_{\g}^{0.357}e^{-\frac{32}{T_{\g}}}$ ${\rm m^{3} s^{-1}}$ given in Ref. \citep{2012Pritchard}. 
The redshift evolution of the gas kinetic temperature can be described by\citep{2007Khatri}
\begin{equation} \label{tg}
\frac{\partial T_{\g}}{\partial z}=\frac{2T_{\g}}{1+z}-\frac{32\sigma_{\rm T} T_{o}^{4}\sigma_{\rm SB}}{3m_{e}c^{2} H_{o}\sqrt{\Omega_{\rm m0} }} (T_{\gamma}-T_{\g})(1+z)^{3/2} \frac{x}{1+x},
\end{equation}
where $\sigma_{\rm T}$,  $\sigma_{\rm SB}$ and $x=n_{e}/n_{\rm H}$ are the Thomson scattering cross section, Stefan-Boltzmann constant, and  fractional ionization respectively.  Here, we assume $n_{\rm H}=n_{\HI}+n_{\rm HII}$ and $n_{\rm HII}=n_{e}$, assuming helium to be fully neutral. We see that the evolution of $T_{\rm g}$ is determined by two processes: i) adiabatic cooling of HI gas due to Universe's expansion (first term, rhs ) and ii)  heat flow from the CMBR to gas through its interaction with free electrons (second term, rhs).  Note that $T_{\g}$ depends on the fractional ionization $x$ which, at any redshift, is determined by the interplay between recombination and ionization processes. The ionization fraction $x$ can be calculated at any redshift z before reionization using the equation\citep{1968Peebles}
 \begin{equation} \label{x}
\frac{dx}{dz} = \frac{C}{H(z)(1+z)} \Big[ \alpha_{e} x^{2} n_{\rm H} -\beta_{e}(1-x) e^{-h_{p}  \nu_{\alpha}/k_{B} T_{\g}} \Big],
\end{equation}
where $C = \frac{1+ K \Lambda (1-x) n_{H}}{1+K(\Lambda+\beta_{e})(1-x) n_{H} }$ is the probability of a  hydrogen atom at the first excited state jumping to the ground state without exciting an adjacent ground state atom, $\Lambda=8.3 \, {\rm s}^{-1}$ is the two photon 2s $\rightarrow$ 1s transition rate,  $K=\frac{\lambda_{\alpha}^{3}}{8\pi H(z)}$ accounts for redshifting of Ly$\alpha$ photons due to Universe's expansion, ${h_{p}  \nu_{\alpha}=10.2 {\rm eV}}$, and $\alpha_{e}$ and $\beta_{e}$ are  the hydrogen recombination and photoionization coefficients respectively. Both quantities $\alpha_{e}$ and $\beta_{e}$ depend on $T_{\g}$, and we use the functional forms given in Ref. \citep{1999Seager,2000Seager}. 
\begin{figure}[h!!!!]
 \includegraphics[width=8.5cm,angle=0]{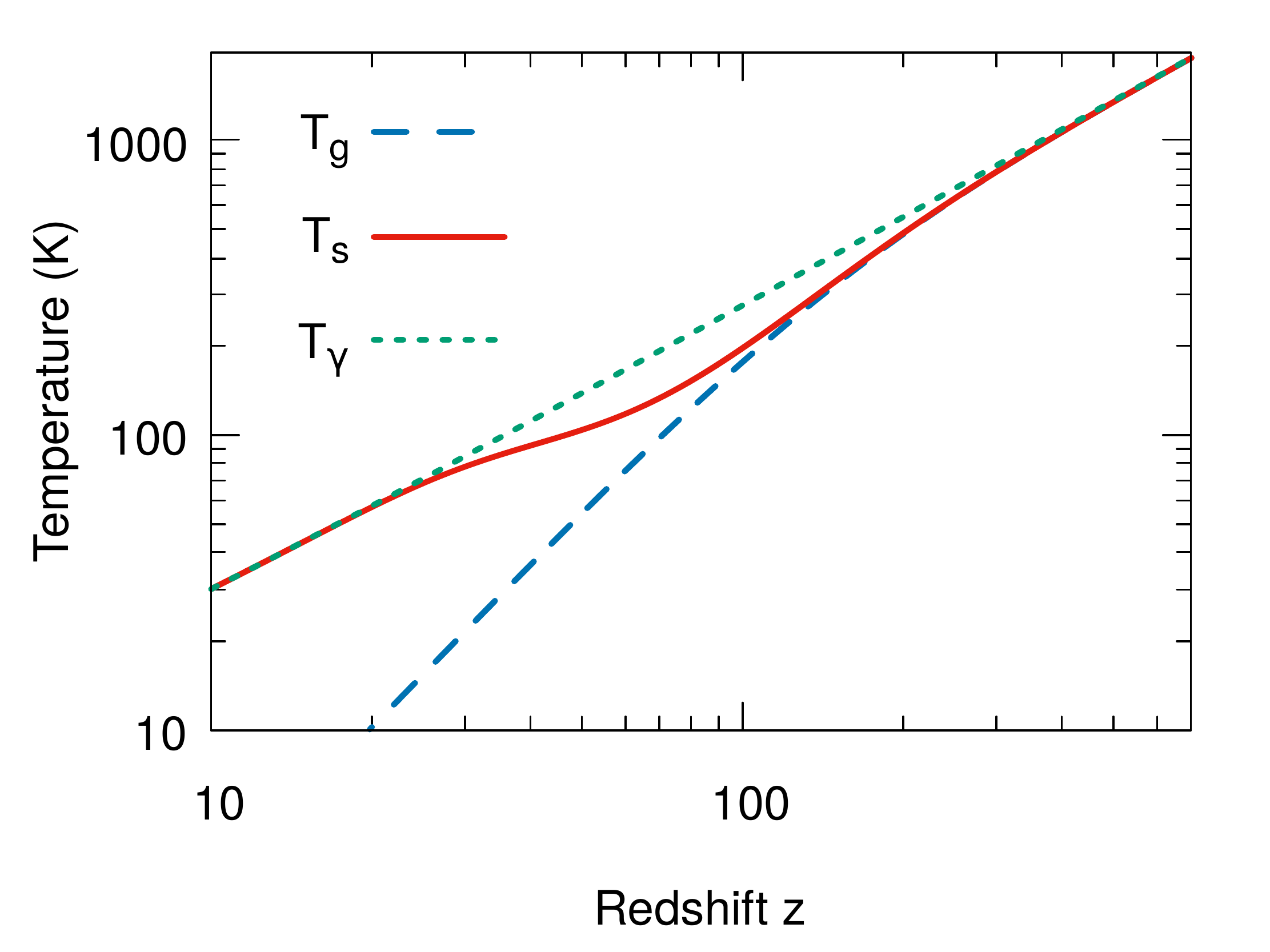}
 \caption{This plot shows the evolution of $T_{\rm g}$, $T_{\rm s}$, and $T_{\gamma}$ with redshift $z$.}
\label{tgtstga}
\end{figure}


Figure \ref{tgtstga} shows the evolution of the CMBR ($T_{\gamma}$), spin ($T_{\rm s}$) and gas kinetic temperature ($T_{\rm g}$) with redshift. Note that the gas kinetic temperature [Eq. \ref{tg}] and ionization equations [Eq. \ref{x}] are coupled and need to be solved together. We compare the solution for $x$ obtained from Eq. \ref{x} with that from the RECFAST code \cite{recfast} and find excellent agreement. We also find that at lower redshifts $z \lesssim 750$, $C$ becomes unity and the photoionization term  becomes negligible compared to the recombination term in the ionization equation (\ref{x}) . As a consequence, the ionization fraction $x$ is essentially determined by the recombination and expansion rate of the Universe at later redshifts.  Heating of gas due to transfer of heat energy from the CMBR  is very efficient at higher redshifts due to higher CMBR temperature and free electron (or gas) density. This keeps $T_{\rm s}$  coupled to $T_{\gamma}$ at redshifts $z \gtrsim 300$. As the Universe expands adiabatically, both the free electron density and the CMBR temperature decrease,  and the CMBR heating of gas becomes less effective in maintaining the gas temperature close to the CMBR temperature. Consequently, $T_{\g}$ starts to decouple from the CMBR and falls below $T_{\gamma}$ at  redshifts  $\lesssim 300$ and eventually scales as $(1+z)^{2}$ at redshifts  $z \lesssim 30$. It means that the effects of the CMBR heating of gas remains up to redshifts $z \sim 30$. The spin temperature $T_{\s}$ remains coupled to $T_{\g}$ up to redshifts $z \gtrsim 100 $ through HI-HI, HI-electron, and HI-proton collisions with the first one dominating over other two during the redshift range of our interest ($30 \lesssim z \lesssim 300$). As the HI gas density and  $T_{\g}$ decrease collisions gradually become inefficient. Meanwhile the radiative coupling of $T_{\s}$ with $T_{\gamma}$ takes over at later redshifts. Consequently, $T_{\s}$ starts to decouple from $T_{\g}$ at redshift $z \lesssim 100 $  and tends toward the CMBR temperature.



\newsec{Fluctuations in differential brightness temperature and impact of inhomogeneous CMBR heating}
Radio interferometric experiments are sensitive to the spatial fluctuations in the 21-cm differential brightness temperature.  The linear perturbation theory is adequate to quantify the fluctuations in the 21-cm signal at large scales during dark ages. It can be seen from Eq. (\ref{tb}) that the fluctuations in the HI differential brightness temperature are a combination of fluctuations in baryon density, spin, and CMBR temperature.  The spin temperature fluctuations arise mainly due to the fluctuations in the gas kinetic temperature and collisional coupling $x_c$.

The fluctuation in baryon density is essentially the same as that in hydrogen density, $\delta_{\rm H}(\vecr,z)=\frac{n_{\rm H}(\vecr,z)-\bar{n}_{\rm H}(z)}{\bar{n}_{\rm H}(z)}$. Fluctuations in $T_{\g}$, $x$, and $T_{\s}$ are defined as $T_{\g}(\vecr,z)={\bar{T}_{\g}(z)}[1+\delta_{\g}(\vecr,z)]$, $x(\vecr, z)={\bar{x}}[1+\delta_{x}(\vecr, z)]$, and $T_{\s}(\vecr,z)={\bar{T}_{\s}(z)}[1+\delta_{\s}(\vecr,z)]$ respectively. The hydrogen density fluctuations, which trace the underlying dark matter (DM) density with a bias factor, essentially lead to fluctuations in all the above quantities. Thus we relate  fluctuations in these quantities to the hydrogen density fluctuations as  $\delta_{\g}=g(\vecr,z)\delta_{\rm H}$, $\delta_{x}=m(\vecr,z)\delta_{\rm H}$, and $\delta_{\s}=s(\vecr,z)\delta_{\rm H}$. Note that all $\delta$s, $g$, $m$, and $s$ are, in general, functions of spatial coordinates $\vecr$ and redshift $z$ but we do not show that explicitly hereafter. Using Eqs. (\ref{tg}) and (\ref{x}) we obtain equations for calculating redshift evolution parameters of linear fluctuations in gas kinetic temperature ($g$) and ionization fraction ($m$) as
\begin{equation} \label{g}
\begin{aligned}
\frac{\partial g}{\partial z}=\left(\frac{2}{3}-g\right)\frac{1}{\delta_{\rm H}}\frac{\partial \delta_{\rm H}}{\partial z}+ \frac{32\sigma_{T} T_{o}^{4}\sigma_{SB}}{3m_{e}c^{2} H_{o}\sqrt{\Omega_{m0} }}(1+z)^{3/2}\times \\
\frac{\bar{x}}{1+\bar{x}}\left[\frac{T_{\gamma}}{\bar{T_{g}}}g-m\left(\frac{T_{\gamma}}{\bar{T_{g}}}-1 \right)\right]
\end{aligned}
\end{equation} 
and, 
\begin{equation} \label{m}
\begin{aligned}
\frac{\partial m}{\partial z}=-\frac{m}{\delta_{\rm H}}\frac{\partial \delta_{\rm H}}{\partial z}+\frac{\bar{C}}{H(z)(1+z)}\big[\bar{\alpha_{e}}\bar{n_{H}}\bar{x} \times \\
\left(\frac{\delta_{C}+\delta_{\alpha}}{\delta_{\rm H}}+1+m \right)-\bar{\beta_{e}}e^{-h_{p}\nu_{\alpha}/k_{B}\bar{T_{g}}} \times \\
 \left\lbrace \frac{m}{\bar{x}}+\frac{(1-\bar{x})}{\bar{x}}\left(\frac{\delta_{C}+\delta_{\beta}}{\delta_{\rm H}}+g\frac{h_{p}\nu_{\alpha}}{k_{B}\bar{T_{g}}} \right) \right\rbrace \big]. 
\end{aligned}
\end{equation}
$\delta_i$ is the first order perturbation term in the $i$th quantity $(i=\alpha , \beta , C)$, where $\delta_{\alpha}=\frac{\partial \alpha_{e}}{\partial T_{g}}\frac{\bar{T_{g}}}{\alpha_{e}(\bar{T}_{g})} \delta_{g}$, $\delta_{\beta}=\frac{\partial \beta_{e}}{\partial T_{g}}\frac{\bar{T_{g}}}{\beta_{e}(\bar{T}_{g})}\delta_{g}$ and $\delta_{C}=\frac{1}{\bar{C}}\Big(\frac{\partial C}{\partial T_{g}}\bar{T_{g}}g+\frac{\partial C}{\partial x}\bar{x}m+\frac{\partial C}{\partial n_{H}}\bar{n_{H}}\Big)\delta_{\rm H}$. Note that Eq. (\ref{g}) is same as the Eq. (25) in Ref. \cite{2014Haimoud}  and Eq. (6) in Ref. \cite{Lewis2007} but represented differently. Equation \ref{m} is equivalent to the Eq. (30) of Ref. \cite{2014Haimoud} and Eq. (3) of Ref. \cite{Lewis2007} if we neglect the photoionization term $\beta_e$ and take $C=1$, which are true for low redshifts. We ignore fluctuations in the CMBR temperature as it is found to be smaller compared to others in the redshift range of our interest.  We note that the quantity  $\delta_{\rm H}$  traces the dark matter density distribution at large scales and lower redshifts. In these situations, the quantity $\frac{1}{\delta_{\rm H}} \frac{\partial \delta_{\rm H}}{\partial z}$ reduces to $\frac{-1}{1+z}$. However, we do not make any such assumption and explicitly calculate the quantity  $\frac{1}{\delta_{\rm H}} \frac{\partial \delta_{\rm H}}{\partial z}$  for different $k$-modes.

We use Eq. (\ref{ts}) and find the redshift evolution parameter of linear fluctuations in the spin temperature as
\begin{equation} \label{s}
s=\frac{x_{c}}{1+x_{c}} \left[ g\frac{\bar{T_{s}}}{\bar{T_{g}}}- \left( \frac{\bar{T_{s}}}{\bar{T_{g}}}-1\right)\left\lbrace1+g\frac{d( \ln K^{H H}_{10})}{d(\ln T_{g})} \right\rbrace\right]. 
\end{equation}
We note that Eqs. (\ref{g}) and (\ref{m}) are coupled; i.e, the fluctuations in the fractional ionization and the kinetic temperature influence each other and eventually affect the fluctuations in spin temperature. This is due to the fact that the heating rate increases with the ionization fraction. Consequently, places with higher ionisation fraction will have higher kinetic temperature.

The spatial fluctuations in the HI differential brightness temperature, defined as $T_b(\vecr,z)= \bar{ T_b}(z)+{\delta T_{b}} (\vecr,z) $, are shown to be directly dependent on the fluctuations in hydrogen density $\delta_{\rm H}$, dark matter density $\delta_{\rm DM }$, and spin temperature as \citep{2004Bharadwaj}
\begin{equation} \label{diffB}
{\delta T_b}(\vecr,z)= \bar{ T}(z)\Big[ \Big( 1-\frac{T_{\gamma}}{\bar{T}_{s}}+ \frac{T_{\gamma}}{\bar{T}_{s}}s\Big) \delta_{\rm H} +\mu^{2}\Big( 1-\frac{T_{\gamma}}{\bar{T}_{s}} \Big)\delta_{\rm DM} \Big], 
\end{equation}
where $\mu=\cos \theta $ and $\theta$ is the angle between the line of sight and $k$-mode. The second term inside the third brackets arises due the peculiar velocity of HI gas. In the above equation, we neglect the fluctuations in $x_{\rm HI}$ as including it shall result in the addition of a term proportional to $\bar{x}m(z)\left(1-\frac{T_{\gamma}}{\bar{T}_{s}} \right)$ which is $O(10^{-5})$ in the redshift range of interest. The HI 21-cm power spectrum $P_{T_{b}}(k, z)$ can be directly linked to the dark matter matter power spectrum $P(k,z)$ as follows, 
\begin{equation} \label{powerspec}
\begin{aligned}
P_{T_{b}}(k,z)=\bar{T}^2(z)\Big[\Big( \frac{T_{\gamma}}{\bar{T}_{s}}-s\frac{T_{\gamma}}{\bar{T}_{s}}-1 \Big)b(k, z) \\
+\mu^{2}\Big( \frac{T_{\gamma}}{\bar{T}_{s}}-1 \Big)\Big]^2P(k,z)
\end{aligned}
\end{equation} 
where $\delta_{\rm H}=b(k, z)\delta_{\rm DM}$. $b(k, z)$ significantly differs from unity at redshift range and scales of our interest \cite{2007naoz, 2007Lewis} and impacts our results.  This quantity has been calculated using CLASS software \citep{class,class2}. Apart from the mean quantities and the underlying DM power spectrum, the HI 21-cm power spectrum depends on fluctuations in the spin temperature through \textbf{s}, which is related to fluctuations in the kinetic temperature and coupling coefficient [see Eq. (\ref{s})]. This tells us that fluctuations in the kinetic temperature $g$, which is caused by inhomogeneous CMBR heating of gas, ultimately affect the HI 21-cm power spectrum.  We discuss this in detail in the subsequent section. 

\begin{figure}[h!!!!!]
 \includegraphics[width=9.0cm,angle=0]{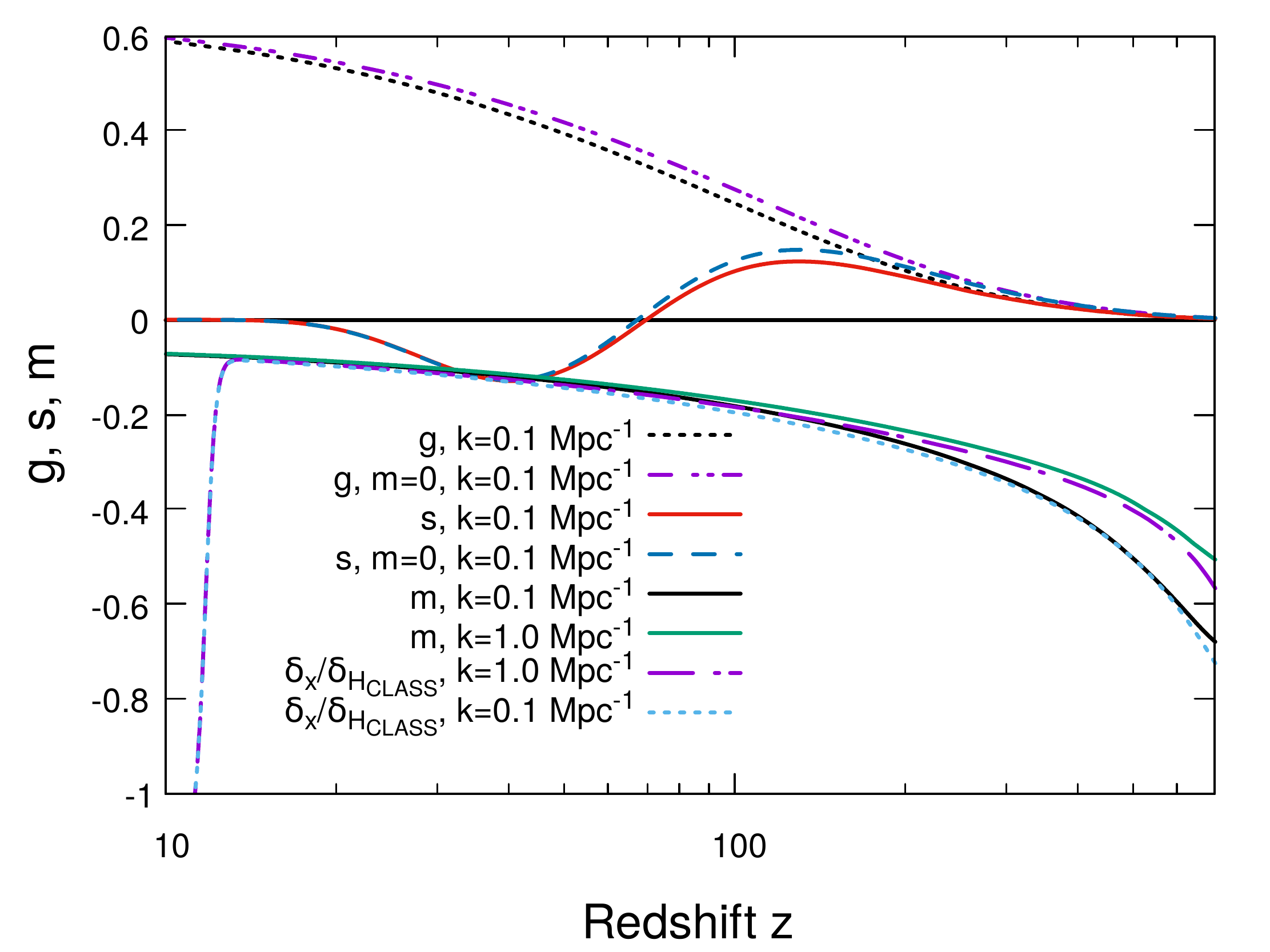}
 \caption{This plot shows the redshift evolution of $m(z)$, $g(z)$, and $s(z)$ in situations when the effect of the inhomogeneous CMBR heating of gas is considered [$m (z) \neq 0$] and not considered [$m (z) = 0$].  $m(z)$ has been shown for two $k$-modes. The quantity $\delta_{\rm x}/\delta_{{\rm H}_{\rm CLASS} }$, which is essentially the same as $m(z)$, has been calculated using CLASS software.}
 \label{gsmgs}
 \end{figure}
 
  \begin{figure}[h] 
\includegraphics[width=9.0cm,angle=0]{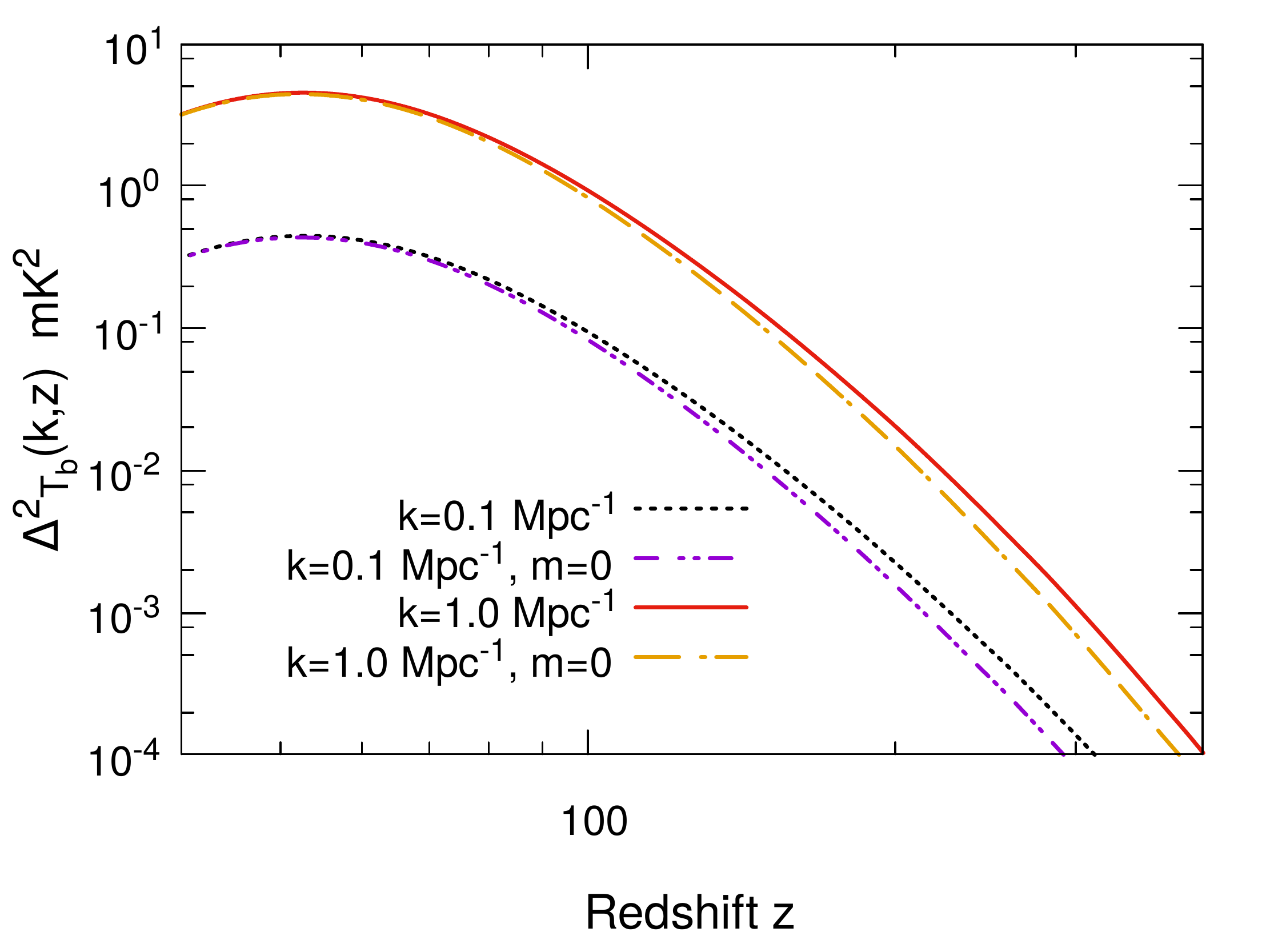}
\caption{This plot shows redshift evolution of dimensionless HI 21-cm power spectrum $\Delta^2_{Tb}(k,z)=k^3 \times P_{Tb}(k,z)/{2 \pi^2}$ for two $k$-modes for $m=0$ (no inhomogeneous CMBR heating) and $m \neq 0$. }  
\label{power}
\end{figure}

\newsec{Results}
\label{sec:results}
Figure \ref{gsmgs} shows the redshift evolution of fluctuations in the kinetic temperature ($g$), spin temperature ($s$) at $k=0.1 \hspace{0.1cm} {\rm Mpc}^{-1}$ and fluctuations in ionization fraction ($m$) at $k=0.1 \hspace{0.1cm} {\rm Mpc}^{-1}$ and $k=1.0 \hspace{0.1cm} {\rm Mpc}^{-1}$ respectively. These quantities are set to zero at very high redshift as they are highly coupled to the CMBR and thus expected to have negligible initial fluctuations.  In order to solve Eqs. (\ref{g}) and (\ref{m}), we  first obtain $\delta_{\rm H}$ as a function of redshift z at different $k$-modes from the CLASS software \citep{class,class2} and then compute $\frac{1}{\delta_{\rm H}(k,z)} \frac{\partial \delta_{\rm H}(k, z)}{\partial z}$, which is essentially the same as $\frac{1}{\delta_{\rm H}(r,z)} \frac{\partial \delta_{\rm H} (r, z)}{\partial z}$. Initially, at higher redshifts, the photoionization plays a major role in determining the ionization fraction $x$ as the CMBR temperature is higher and coupled to the gas temperature. However, the recombination process starts dominating at $z \lesssim 1200$ over the photoionization, and, therefore, $m \rightarrow -1$.  This means places with higher gas density have lower ionization fraction because the recombination rate, which is proportional to the gas density, is higher there. As the redshift further decreases, expansion of the Universe plays a significant role in determining $m$, and as a consequence, $m \rightarrow 0$ at later redshifts.   Figure \ref{gsmgs} also compares the evolution of $m$ obtained directly from CLASS \cite{class} software (by taking the ratio of fluctuations in ionization fraction  $\delta_x$ and baryon density $\delta_{\rm H}$) with the same calculated using Eq. (\ref{m}) for two $k$-modes $0.1$ and $1.0 \, {\rm Mpc}^{-1}$,  and they agree with each other quite well.   Reionization due to stellar sources starts at  redshift $z \sim 15$ in the model considered in  CLASS which leads to the deviation in $m$ observed at redshift $z \lesssim 15$.

\begin{figure}[h]
\includegraphics[width=6.0cm,angle=270]{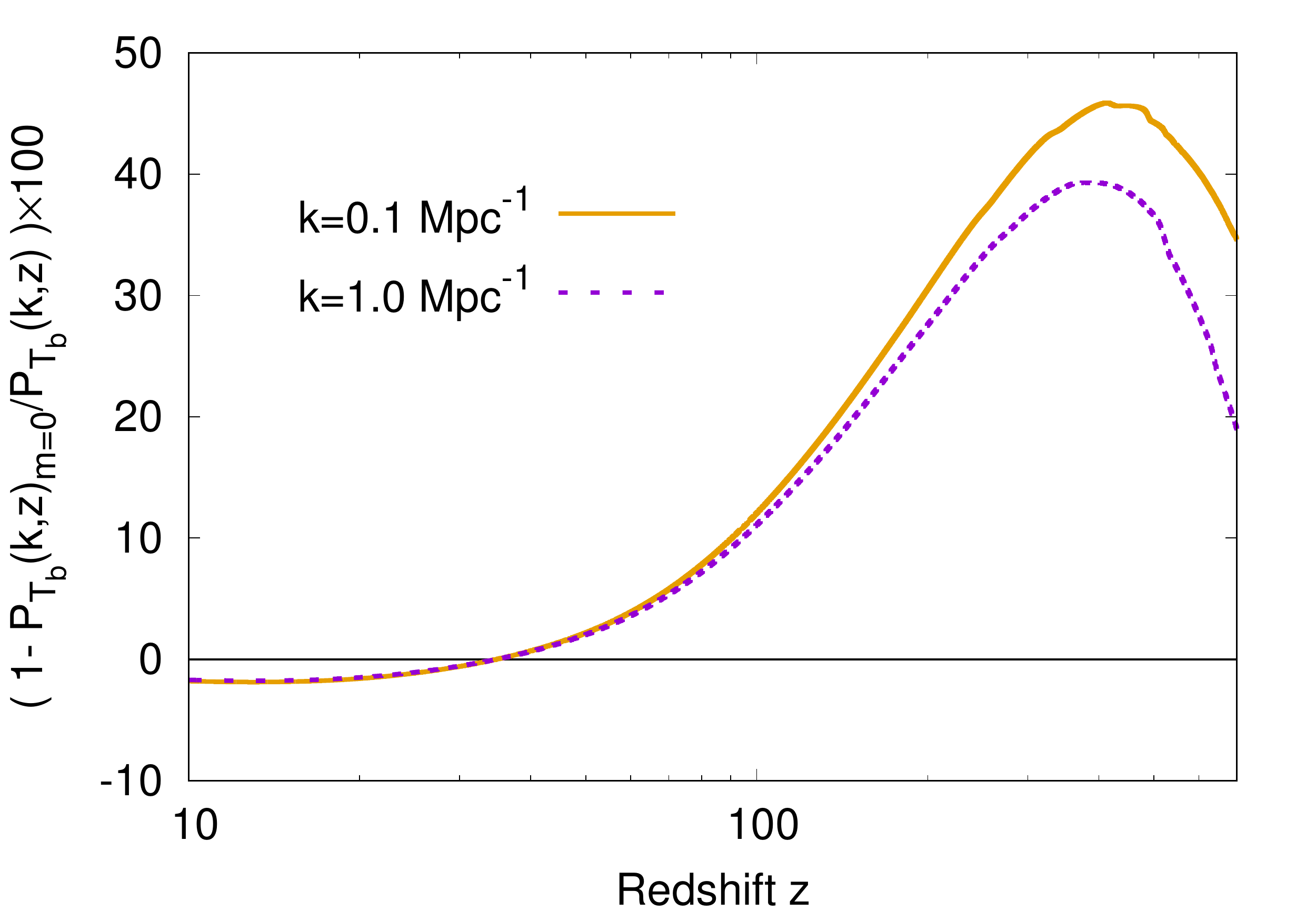}
\caption{This plot shows redshift evolution of the percentage change in the HI 21-cm power spectrum due to inhomogeneous CMBR heating for two $k$-modes and $\mu^2=0.5$.}
\label{rpower}
\end{figure}

Next, we find that $g$, the redshift evolution parameter of linear fluctuations in the gas kinetic temperature, is considerably impacted by the inclusion of fluctuations in the ionization fraction, particularly at redshifts $z \lesssim 300$. In general, $g$ is effectively zero at redshifts $z \gtrsim 300$ when $T_{\rm g}$ is highly coupled to $T_{\gamma}$. It starts to increase from redshift $z \sim 300$ and tends to a value $2/3$ as redshift decreases---a condition for a gas expanding purely adiabatically. However, as we discussed above, the ionization fraction is inhomogeneously distributed. As a result, the CMBR heating of gas, which depends on ionization fraction $x$ [see Eq. (\ref{x})], becomes inhomogeneous too. Places with higher gas density, which have a lower ionization fraction, are less efficient in  transferring heat from the CMBR to gas. Consequently, higher density places, which normally have higher gas kinetic temperatures ($g >0$), will be negatively impacted by the inclusion of the inhomogeneous CMBR heating caused by inhomogeneous ionization fraction. This fact is manifested in Eq. (\ref{g}), which contains $-m(T_{\gamma}/T_{\rm g}-1)$  inside the second term on the rhs. For this reason, $g$ gets lowered when the effects of inhomogeneous CMBR  heating of gas are considered. Obviously, the effect is negligible at higher redshifts as $T_{\rm g}$ and $T_{\gamma}$ are coupled. It becomes important when $T_{\rm g}$ starts to decouple from the CMBR and $g$ grows separately at redshift $z \sim 300$. It continues to be important up to redshifts at which the CMBR heating of gas still plays a significant role in determining $T_{\rm g}$. In other words, the inhomogeneous CMBR heating of gas is an important contributor to the gas kinetic temperature fluctuations during the transition period of $T_{\rm g}$ from being fully coupled to the CMBR to fully decoupled from it.  The other effect that could, in principle, affect the evolution of $g$, $m$ and all subsequent results is the spatial fluctuations in the CMBR density. We consider this effect and find it to have a negligible impact on both quantities  in the redshift range of our interest.

The redshift evolution parameter of linear fluctuations in the spin temperature $s$ closely follows $g$ at redshifts $z \gtrsim 200$ and, therefore, gets suppressed when the effect of the inhomogeneous CMBR heating of gas is included. This is a consequence of the fact that the spin temperature $T_{\rm s}$ including its fluctuations remains coupled with $T_{\rm g}$ through collisional coupling at high redshifts.   This can also be understood from Eq. (\ref{s}). Thereafter, $s$ starts to unfollow $g$ as collisions, which depend on $T_{\rm g}$ and HI density, become less efficient.  Later, $s$ becomes negative and eventually tends to zero at lower redshifts. We note that $s$ gets suppressed in the redshifts range $30 \lesssim z \lesssim 200$. Although the generic nature of the redshift evolution of $m$, $g$, and $s$ remains the same for different $k$-modes,  we find them to mildly increase with $k$-modes in the redshift range of our interest.

We finally investigate the effect of the inhomogeneous CMBR heating of gas on the power spectrum of  HI 21-cm brightness temperature fluctuations. From Eq. (\ref{powerspec}) we see that $s$ is directly linked to the power spectrum in a way in which a suppression of the spin temperature fluctuations leads to an increment in the power spectrum.  Figure. \ref{power} plots the dimensionless HI 21-cm power spectrum $\Delta^2_{Tb}(k,z)=k^3 \times P_{Tb}(k,z)/{2 \pi^2}$ for two $k$-modes  for the case when the  inhomogeneous CMB heating  is considered in the evaluation of $s$ and compares with the case when the effect is not considered, i.e., $m=0$. We assume $\mu^2$ to be $\sim 0.5$. We obtain a considerable enhancement in the HI brightness temperature power spectrum $P_{T_{b}}(k, z)$ for the first case.  At the redshift $z \sim 50$,  the enhancement in the HI power spectrum is $\sim 2 \%$. This is in good agreement with the results reported in Refs. [\cite{2007Lewis,Lewis2007}]\footnote{We note that Refs. [\cite{2007Lewis,Lewis2007}] looked at the effects on the angular power spectrum $C_l$ where as we focus on $P_{Tb}(k,z)$. These two different statistical quantities can be compared only for an \quotes{average} value of $\mu^2$, which is assumed to be $0.5$.}. Further, we find that the HI power spectrum $P_{T_b}(k, z)$ is enhanced by $\sim 4 \%$, $\sim 10 \%$ , $\sim 20 \%$, and $\sim 30 \%$ at redshifts $60$, $90$, $140$, and $200$ respectively at $k=0.1 \, {\rm Mpc}^{-1}$ when  the inhomogeneous CMB heating of gas is considered (see Figure. \ref{rpower}).  We also notice  that this enhancement has a weak dependence on $k$-modes. Although the percentage change is even greater  at higher redshifts, the absolute change in the power spectrum i.e, the quantity ($P_{T_b}-P_{T_b, m=0}$) becomes insignificant  as $T_{\rm s}$ becomes coupled to $T_{\gamma}$. The enhancement in the power spectrum can be explained using Eq. (\ref{powerspec}), which shows the power spectrum's dependence on $s$. As we discussed above, $s$ gets suppressed when the effect of the inhomogeneous CMBR heating of gas is included.

Although the  above paragraph discussed results for  $\mu^2=0.5$, the effect on the power spectrum $P_{Tb}(k, z)$ would be different for different values of $\mu^2$ [see in Eq. (\ref{powerspec})] which varies from $0$ to $1$. For higher values of $\mu^2$, the contribution from the peculiar velocity effect, which is insensitive to the inhomogeneous heating, becomes higher. This reduces the overall effect of the inhomogeneous heating on the HI power spectrum $P_{Tb}(k,z)$. For $\mu^2=1$, we find that $P_{Tb}(k,z)$ is enhanced just by $\sim1.7\%$, $\sim 2.7\%$, $\sim6.4\%$, $11.5\%$, and $\sim16\%$ at redshifts around $50$, $60$, $90$, $140$, and $200$ respectively at $k = 0.1 \, {\rm Mpc}^{-1}$.  On the other hand, the changes are considerably higher at $\sim 3\%$, $\sim 5\%$, $\sim 17.5\%$, $49\%$, and $\sim93\%$  at redshifts around $50$, $60$, $90$, $140$, and $200$ respectively at $k = 0.1 \, {\rm Mpc}^{-1}$ for $\mu^2=0$. However, we note that, the observed signal at modes corresponding to $\mu^2 \lesssim \frac{C^2}{C^2+1}$ will be highly dominated by contributions from strong foreground sources.  Here $C$ is a function of both the antenna primary field of view and redshift $z$  \citep{2014dillon}. Thus, these modes may be avoided while extracting the HI 21-cm power spectrum from observed data. On the other hand, some small $k_{\perp}$-modes are not available in radio interferometric observations due to a lower baseline cut-off. This poses restrictions on using modes corresponding to $\mu^2 \sim 1$.  Consequently, the intermediate modes centered around $\mu^2 \sim 0.5$ are likely to be more useful for HI 21-cm power spectrum measurements.

\newsec{Summary and Discussion}
Observations of the redshifted 21-cm signal seem be the only viable probe of cosmic dark ages. The kinetic temperature of the intergalactic neutral gas, which plays a major role in determining the HI 21-cm signal, is significantly affected during dark ages by the heat transfer from the CMBR to gas through its interaction with free electrons. The heat transfer from the CMBR to gas in the intergalactic medium depends on the ionization fraction, which is inhomogeneously distributed in  space mainly due to the inhomogeneous recombination process. We investigate, in detail, the effect of the inhomogeneous CMBR heating of gas  on the HI 21-cm differential brightness temperature fluctuations over a large redshift range during dark ages. We follow a simple analytical formalism which clearly explains roles of relevant physical processes leading to the effect. We find that the inhomogeneous heating of gas causes additional fluctuations in the kinetic temperature and, consequently, in the spin temperature and ultimately in the 21-cm signal.  We also find that the effect has detectable signatures in the HI 21-cm power spectrum when the gas kinetic temperature $T_{\rm g}$ starts to decouple from the CMBR  at redshift $z \sim 300$.  The effect remains important down to redshift $z \sim 30$ up to which the CMBR heating of gas plays a significant role in determining $T_{\rm g}$.  In other words, the transition period of $T_{\rm g}$ being fully coupled to the CMBR to fully decoupled from it remains important for the effect. Our results agree quite well with earlier studies which focus only on redshift $\sim 50$ and find a negligible effect on the HI power spectrum.  However, we find that the HI power spectrum $P_{T_b}(k, z)$ is considerably enhanced by $\sim 4 \%$, $\sim 10 \%$ , $\sim 20 \%$ and $\sim 30 \%$ at redshifts $60$, $90$, $140$, and $200$ respectively for $k=0.1 \, {\rm Mpc}^{-1}$ and $\mu^2 =0.5$. The effect becomes even higher for lower values of $\mu^2$ due to the reduced influence of the peculiar velocity which is insensitive to the inhomogeneous heating. We also notice  a mild change in our results at higher $k$-modes. At higher redshifts $z \gtrsim 300$ the effect makes negligible changes in observable quantities as the spin temperature gets coupled to the CMBR temperature.  The effect has its root in the underlying matter and gas density fluctuations and, therefore, is intrinsic to any calculation regarding HI 21-cm signal during dark ages. It has to be considered in every situation where the gas kinetic and spin temperature fluctuations play important roles.



\smallskip
{\it Acknowledgements: K.K.D. and D.D.C. acknowledge financial support through  DST project SR/FTP/PS-119/2012. We thank Sk. Saiyad Ali for useful discussion and his help. We also thank CTS, IIT Kharagpur for a visit during which a part of this work was done.} 

\end{document}